\documentclass[%
 aip,
 amsmath,amssymb,
 reprint,%
]{revtex4-1}

\usepackage{graphicx}
\usepackage{dcolumn}
\usepackage{bm}
\usepackage{float}
\usepackage[utf8]{inputenc}
\usepackage[T1]{fontenc}
\usepackage{mathptmx}
\usepackage{etoolbox}
\usepackage{newunicodechar}
\newunicodechar{（}{(}
\newunicodechar{）}{)}

\usepackage[colorlinks=true,
            linkcolor=blue,     
            citecolor=blue,  
            urlcolor=blue]{hyperref} 
\usepackage[capitalize, nameinlink]{cleveref}
\crefname{figure}{Figure}{Figures}
\Crefname{figure}{Figure}{Figures}

\makeatletter
\def\@email#1#2{%
 \endgroup
 \patchcmd{\titleblock@produce}
  {\frontmatter@RRAPformat}
  {\frontmatter@RRAPformat{\produce@RRAP{*#1\href{mailto:#2}{#2}}}\frontmatter@RRAPformat}
  {}{}
}%
\makeatother
\begin{document}

\preprint{AIP/123-QED}

\title{Batch-Fabricated PDMS Templates for the Robotic Transfer of 2D Materials}
\author{Zhili Lin}
 \author{Luosha Han}%
 \author{Jinkun He}
 \author{Xiaoxue Fan}
  \affiliation{State Key Laboratory of Quantum Optics Technologies and Devices, Institute of Optoelectronics, Shanxi University, 92 WuCheng
Road, Taiyuan, 030006, Shanxi, China}
  \affiliation{ 
Collaborative Innovation Center of Extreme Optics, Shanxi University, 92 WuCheng Road, Taiyuan, 030006, Shanxi, China
}
\author{Tongyao Zhang}
\author{Xiaoxi Li}
  \affiliation{State Key Laboratory of Quantum Optics Technologies and Devices, Institute of Optoelectronics, Shanxi University, 92 WuCheng
Road, Taiyuan, 030006, Shanxi, China}
  \affiliation{ 
Collaborative Innovation Center of Extreme Optics, Shanxi University, 92 WuCheng Road, Taiyuan, 030006, Shanxi, China
}
 \affiliation{ 
Liaoning Academy of Materials, 280 Chuangxin Road, Shenyang, 110167, Liaoning, China
}
\author{Baojuan Dong}
  \affiliation{State Key Laboratory of Quantum Optics Technologies and Devices, Institute of Optoelectronics, Shanxi University, 92 WuCheng
Road, Taiyuan, 030006, Shanxi, China}
  \affiliation{ 
Collaborative Innovation Center of Extreme Optics, Shanxi University, 92 WuCheng Road, Taiyuan, 030006, Shanxi, China
}
 \affiliation{ 
Liaoning Academy of Materials, 280 Chuangxin Road, Shenyang, 110167, Liaoning, China
}
\affiliation{ 
Hefei National Laboratory, 5099 Wangjiang West Road, Hefei, 230088, Anhui, China
}
\author{Kai Zhao}
  \affiliation{State Key Laboratory of Quantum Optics Technologies and Devices, Institute of Optoelectronics, Shanxi University, 92 WuCheng
Road, Taiyuan, 030006, Shanxi, China}
  \affiliation{ 
Collaborative Innovation Center of Extreme Optics, Shanxi University, 92 WuCheng Road, Taiyuan, 030006, Shanxi, China
}
 \affiliation{ 
Liaoning Academy of Materials, 280 Chuangxin Road, Shenyang, 110167, Liaoning, China
}
 \email{dongbaojuan.1989@gmail.com, kaizhao@sxu.edu.cn}
%

 \homepage{http://www.Second.institution.edu/~Charlie.Author.}

\date{\today}

\begin{abstract}
Robotic stacking of van der Waals heterostructures has been at the verge thanks to the convergence between artificial intelligence (AI) and two-dimensional (2D) materials research. Key ingredients to fulfill this pursuit often include algorithms to identify layer compounds on chips, hard-wares to realize sophisticated operations of motion and/or rotation in a microscale, and, as importantly, highly-standardized and uniform transfer stamps that are often used in picking up layered materials under a microscope. Here, we report a hot-casted-droplet batch fabrication method for polydimethylsiloxane (PDMS) templates tailored for dry transfer of 2D materials. Controlled precursor formulation, degassing, and motorized-syringe dispensing produce dome-shaped PDMS templates with ultra-smooth surfaces (root-mean-square roughness $\sim$ 0.3 nm at relatively low curing temperatures). By tuning the curing temperature, the reproducible and controllable apex curvature allows precisely defined contact area between the organic adhesive film and substrate, via thermal expansion. Our results further reveals thermalmechanical behaviors with different casting parameters of such PDMS domes. This scalable and parameterized fabrication protocol gives rise to uniform transfer-stamps with ultra-smooth surface, which may be beneficial for future AI-driven robotic assembly of 2D material heterostructures. 
\end{abstract}

\maketitle



\begin{figure}[!htbp]
\label{fig:1target}
	\centering
\includegraphics[width=.95\columnwidth]{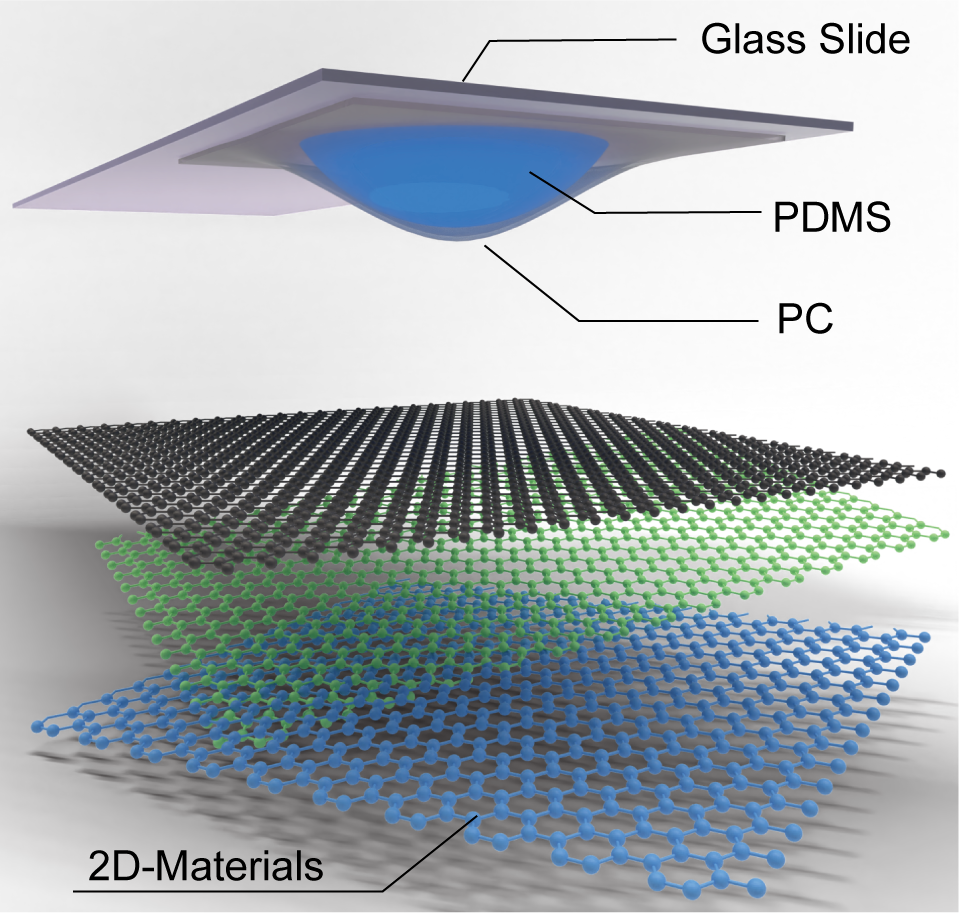}
	\caption{Schematic of the PDMS-based stamp for dry transfer of 2D materials. The stamp consists of a glass slide, a hemispherical PDMS template, and a polycarbonate (PC) film. Vertical motion controlled by a precision mechanical stage allows the PC film to pick up 2D flakes via controlled adhesion. The flexible PDMS template enables gradual, programmable contact with the target substrate, supporting iterative assembly of multilayer van der Waals heterostructures and quantitative AI-assisted robotic transfer.}
	\label{FIG:1}
\end{figure}

Since the discovery of graphene in 2004, two-dimensional (2D) materials have revolutionized condensed matter physics and nanotechnology owing to their exceptional electrical, optical, and mechanical properties.\cite{novoselov2004electric} The capability to vertically assemble 2D layers into van der Waals (vdW) heterostructures has further expanded this field, enabling the exploration of correlated quantum phenomena such as superconductivity, Mott insulating states, and topological excitations in twisted bilayer and multilayer systems.\cite{cao2018unconventional,lu2019superconductors,li2021quantum,wang2022quantum,yang2023unconventional,sha2024observation,xie2025tunable} Beyond fundamental discoveries, these artificially stacked systems hold great promise for applications in quantum devices, nanoelectronics, and optoelectronic integration.\cite{liu2019van,chen2020finfet,guo2024van} To realize such complex architectures, the dry-transfer technique has become the cornerstone of vdW assembly, offering precise control over the stacking sequence, orientation, and alignment of individual atomic layers.\cite{meitl2006transfer,wang2013one,zomer2014fast} In this method, a flexible polydimethylsiloxane (PDMS) stamp—often coated with an organic adhesive film such as polycarbonate (PC) or polypropylene carbonate (PPC)—is used to pick up and release exfoliated flakes under an optical microscope, as shown in \hyperref[FIG:1]{Fig.\@ 1}. The elasticity of the PDMS enables gradual, conformal contact between the adhesive film and the substrate, minimizing mechanical stress on the delicate 2D materials during transfer.

\begin{figure*}[t]
    \phantomsection
    \label{fig:2target}
	\centering
	\includegraphics[width=0.99\textwidth]{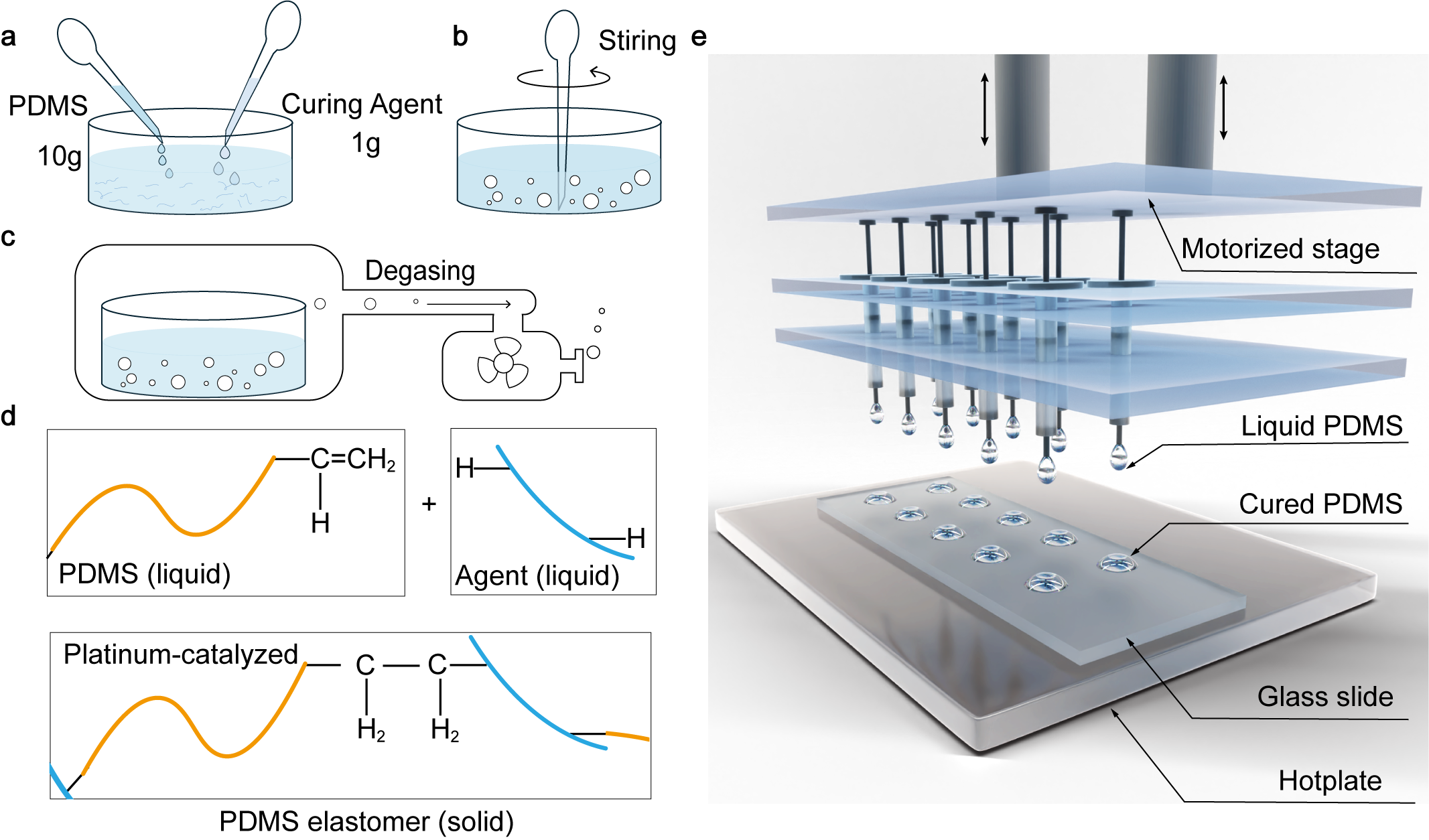}
	\caption{Preparation of the PDMS precursor and crosslinking reaction mechanism. (a) Mixing the base prepolymer and crosslinking agent at a 10:1 weight ratio. (b) Mechanical stirring to ensure homogeneous blending. (c) Vacuum degassing to eliminate entrapped air bubbles. (d) Schematic illustration of the Platinum-catalyzed crosslinking reaction during thermal curing. (e) Schematic of the arrayed microinjector system for precise and reproducible batch dispensing of PDMS templates, the system integrates a motorized stage, an injector array, and a temperature-controlled hotplate, enabling uniform deposition of PDMS droplets onto preheated glass slides for subsequent 2D material transfer.}
	\label{FIG:2}
\end{figure*} 
\begin{figure*}[t]
    \phantomsection
    \label{Fig:3target}
   
	\centering
	\includegraphics[width=0.95\textwidth]{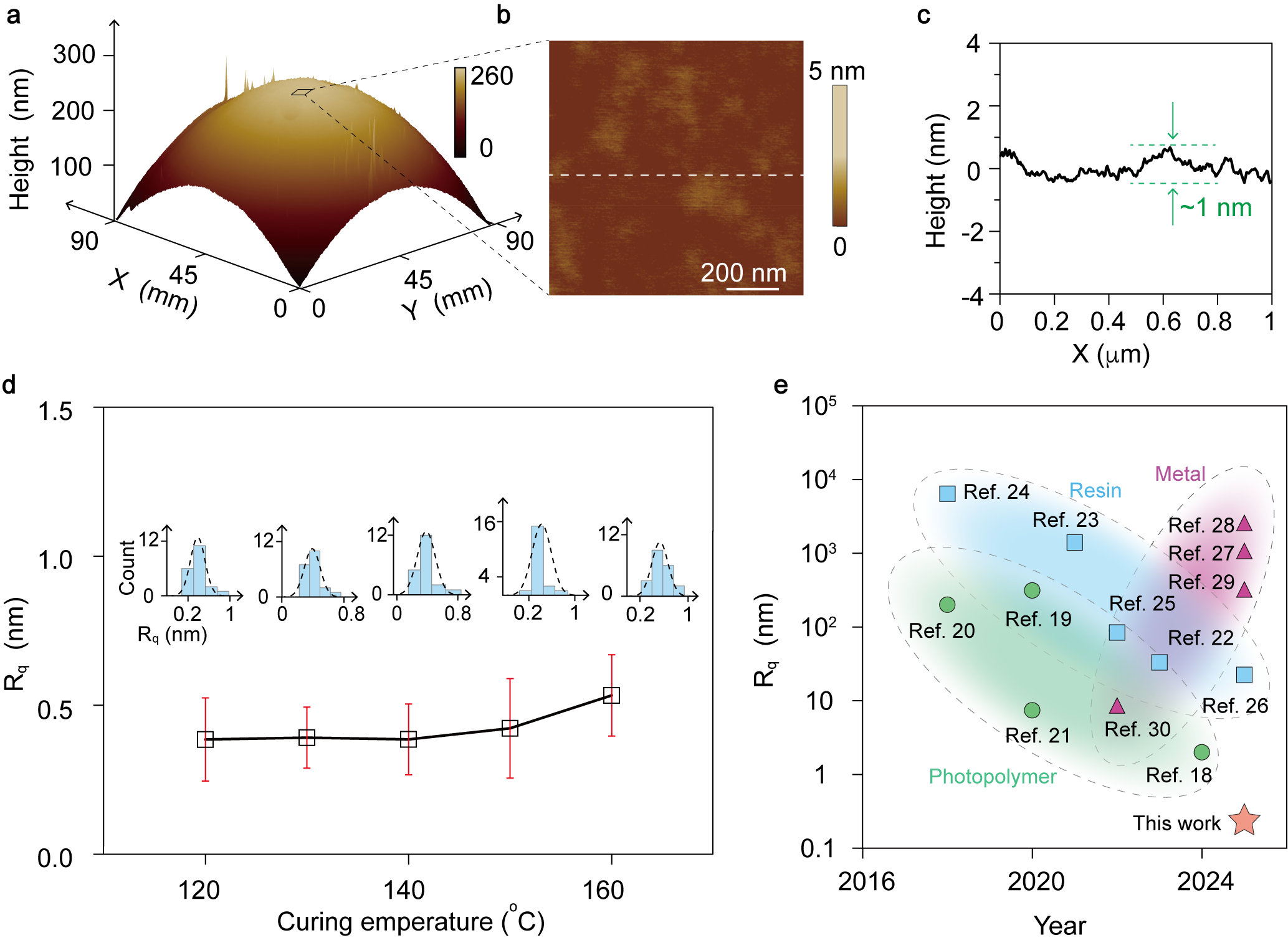}
	\caption{Surface roughness characterization of PDMS templates.  (a) Three-dimensional AFM topography of a typical PDMS template cured at 120 \textcelsius{}. (b) High-resolution AFM height map obtained from a $1~\mu\mathrm{m}^2$ area at the template apex. (c) Height profile extracted along the dashed line in panel (b). (d) Root-mean-square roughness ($R_q$) as a function of curing temperature, with error bars indicating the standard deviation. Insets show the corresponding roughness distribution histograms. (e) Comparison of PDMS surface roughness prepared using metal, resin, and photopolymer molds versus the proposed batch-fabrication method. Dashed lines and color gradients are included as visual guides to distinguish the different fabrication routes.}
	\label{Fig:3}
\end{figure*} 
\begin{figure*}[!t]
    \phantomsection
    \label{fig:4target}
	\centering
	\includegraphics[width=1\textwidth]{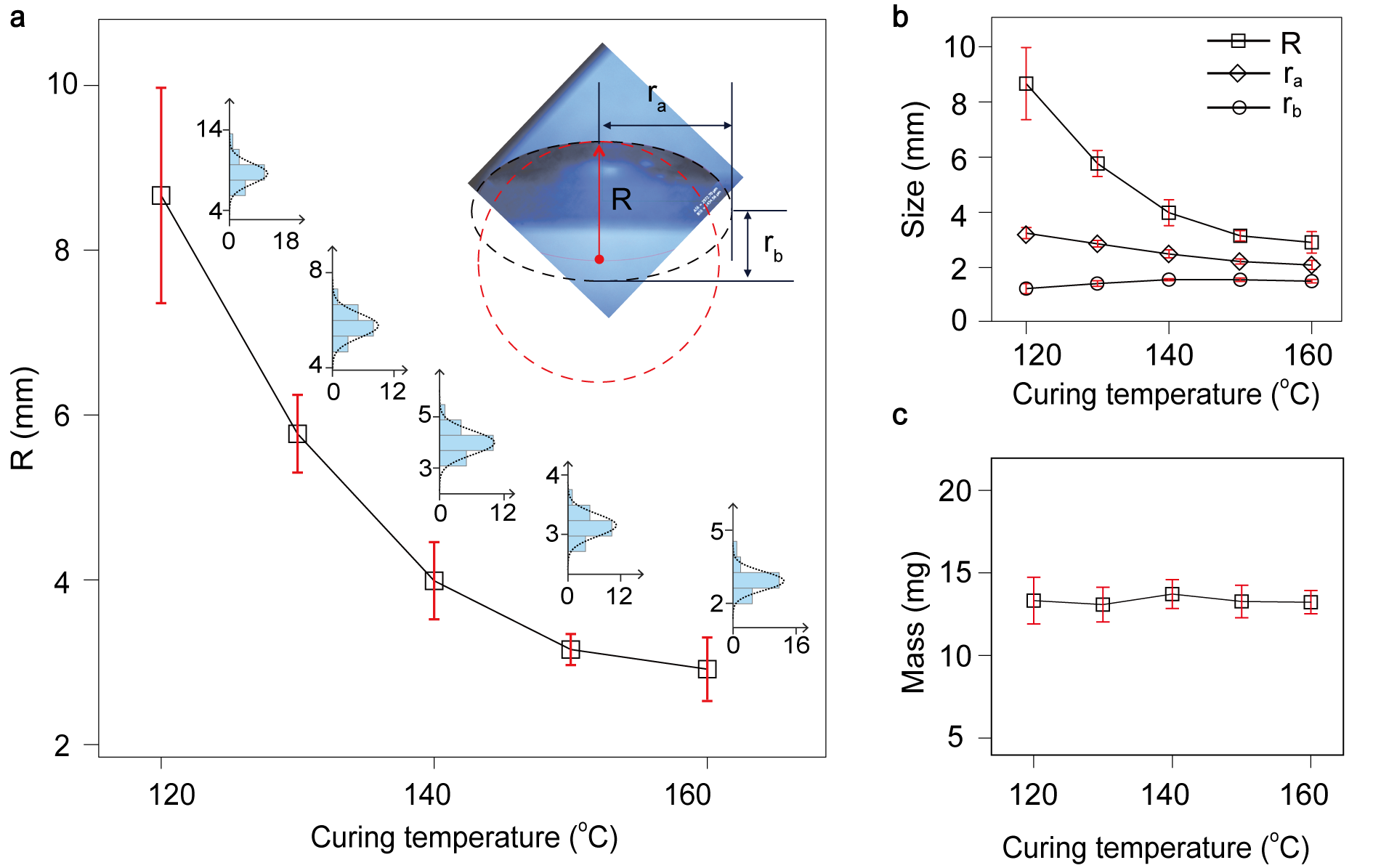}
	\caption{Temperature-dependent geometry and mass of batch-fabricated PDMS templates. (a) Apex curvature radius ($R$) as a function of curing temperature, with error bars representing the standard deviation. The top-right inset shows a representative hemi-ellipsoidal PDMS template and its geometric parameters: $r_a$ (major semi-axis) and $r_b$ (minor semi-axis). Histograms display the distribution of $R$ obtained from 20 replicate templates at each curing temperature. (b) Evolution of $R$, $r_a$, and $r_b$ with curing temperature, illustrating the coordinated dimensional changes induced by accelerated cross-linking dynamics at elevated temperatures. (c) Template mass as a function of curing temperature, demonstrating excellent batch-to-batch reproducibility across all curing conditions.}
	\label{FIG:4}
\end{figure*}
\begin{figure*}[t]
    \phantomsection
    \label{fig:5target}
	\centering	\includegraphics[width=1\textwidth]{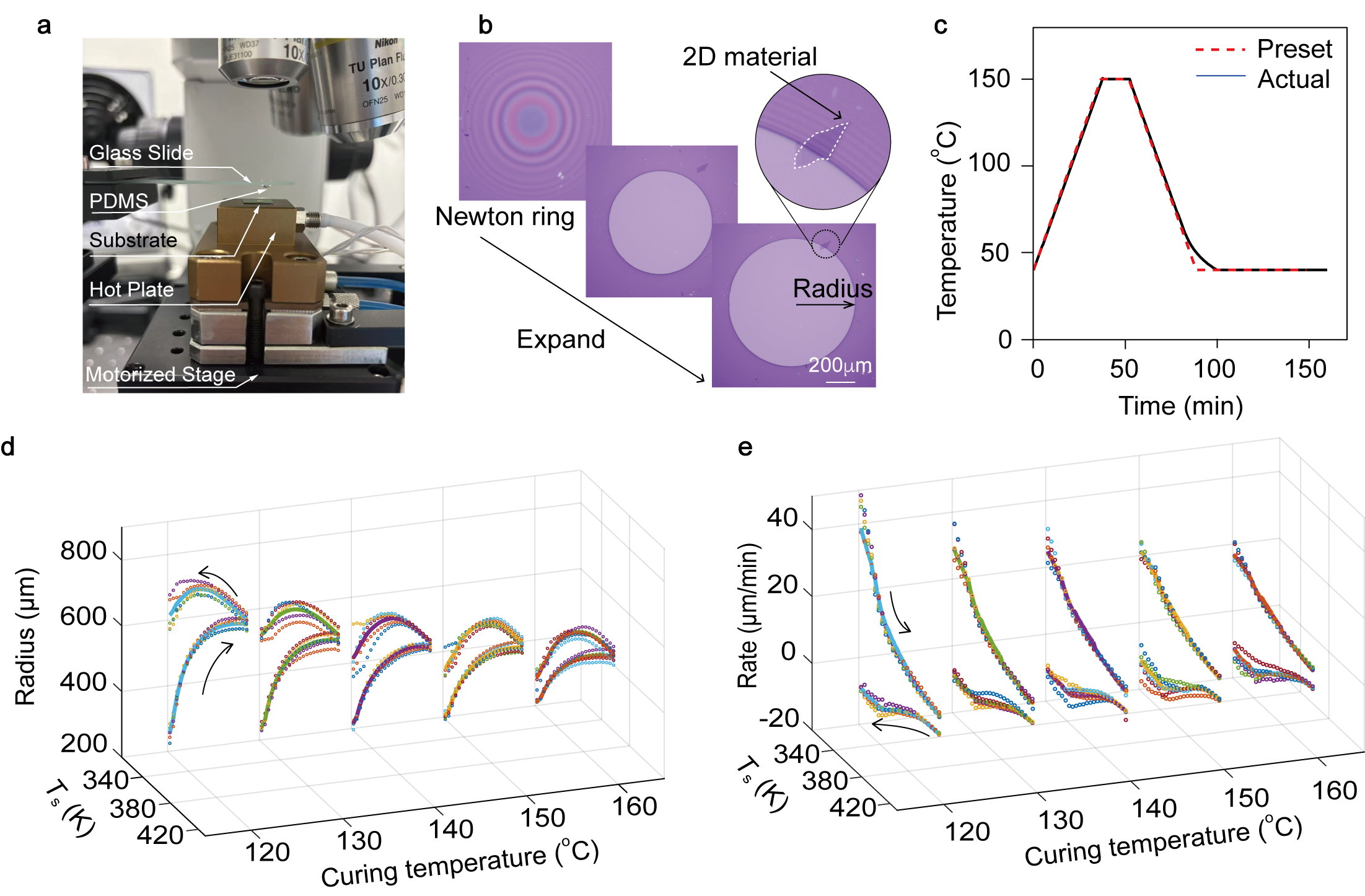}
	\caption{Thermomechanical behavior of PDMS templates at different curing temperatures. (a) Schematic of the dry-transfer setup, including the transfer stamp, motorized precision stage, and hotplate. (b) Optical images showing the formation of initial Newton's rings and the gradual expansion of the circular contact area toward the target 2D material edge. (c) Programmed and measured substrate temperature profiles during the transfer process. (d) Contact radius versus substrate temperature ($T_s$) during thermal cycling for PDMS templates fabricated at different curing temperatures. Forward and backward arrows indicate heating and cooling processes, respectively. (e) Corresponding thermal expansion rate as a function of substrate temperature for the same template sets, with arrows denoting heating/cooling directions.}

	\label{FIG:5}
\end{figure*}
Recently, the convergence of artificial intelligence (AI) and 2D materials research has pushed dry transfer toward robotic automation. AI-assisted optical recognition of exfoliated flakes, closed-loop control of micromechanical motion, and programmable thermal management are gradually transforming the manual, experience-dependent stacking process into a reproducible and intelligent robotic operation. However, to achieve truly autonomous stacking, standardized and highly uniform PDMS transfer templates are essential. The thermomechanical behavior of the PDMS, especially its geometry, surface smoothness, and temperature-dependent expansion, critically determines the contact dynamics and reproducibility of the transfer process. Here, we report a hot-casted-droplet batch fabrication method for dome-shaped PDMS templates tailored for dry transfer of 2D materials. The process integrates controlled precursor formulation, degassing, and motorized-syringe dispensing, yielding PDMS domes with ultra-smooth surfaces (root-mean-square roughness $\sim$ 0.3 nm at lower curing temperatures) and reproducible apex curvature. By tuning the curing temperature, the thermal expansion of the PDMS can be precisely regulated, allowing predictable control of the contact area between the organic adhesive film and the substrate. Furthermore, systematic investigation of thermomechanical behaviors under different curing conditions establishes a parameterized fabrication framework for reproducible, AI-ready transfer stamps. This scalable method lays the groundwork for future AI-driven robotic assembly of 2D material heterostructures with quantitative and predictable physical interfaces.

\hyperref[fig:2target]{Fig.\@ 2} illustrates the preparation workflow of the polydimethylsiloxane (PDMS) precursor. A silicone elastomer kit (Sylgard 184), comprising a PDMS prepolymer base (Part A) and a crosslinking curing agent (Part B), was used. The base and curing agent were mixed at a 10:1 weight ratio [\hyperref[fig:2target]{Fig.\@ 2(a)}] and stirred for 10 minutes to achieve a homogeneous mixture [\hyperref[fig:2target]{Fig.\@ 2(b)}], which is critical for uniform curing and consistent mechanical properties in the final templates.\cite{li2023functional} The mixture was subsequently degassed at room temperature for 1 hour using a molecular vacuum pump [\hyperref[fig:2target]{Fig.\@ 2(c)}]. As illustrated in \hyperref[fig:2target]{Fig.\@ 2(d)}, the liquid-to-solid transformation occurs via a platinum-catalyzed reaction between vinyl-functionalized PDMS and Si–H groups in the curing agent, forming a crosslinked network that ensures reliable elasticity and mechanical.\cite{network1, network2} To achieve precise control of PDMS droplet volume and enable reproducible batch fabrication, a microinjector-assisted dispensing system was employed [\hyperref[fig:2target]{Fig.\@ 2(e)}]. The system comprises a motorized mechanical controller, an injector array, and a temperature-controlled hotplate. The motorized mechanical controller drives the motorized stage to descend according to preset extrusion parameters (plunger speed and duration), enabling steady ejection of PDMS precursor through the injector nozzle. After optimization, the extrusion parameters were set at a plunger speed of 3 $\mu$L/s for 4 s, ensuring clean droplet detachment without residue or filament formation. Each injection cycle deposits a 2 × 5 array of PDMS templates onto a preheated glass slide, with the PDMS and slide maintained on the hotplate for more than 20 s, and sequential repetition enables batch fabrication with high uniformity and reproducibility.

The principle behind achieving uniform and controlled PDMS droplet arrays relies on precise mechanical regulation of plunger displacement and extrusion rate. By synchronizing the descent of the motorized stage with the controlled opening of each injector nozzle, the system ensures that each droplet receives an identical volume of PDMS precursor. The viscoelastic properties of the PDMS mixture, combined with surface tension effects at the nozzle exit, naturally shape the droplet into a reproducible hemispherical profile. This active control over volume and droplet detachment prevents filament formation, coalescence, or positional deviation, producing highly uniform template arrays suitable for subsequent 2D material transfer. During dispensing, the hotplate temperature is a critical parameter that directly determines the apex curvature of the PDMS template, which strongly affects the efficiency and reliability of 2D material transfer. In this work, the curing temperature was confined to 120 – 160 \textcelsius{}, producing an optimal apex curvature suitable for dry transfer of 2D materials.


The surface roughness of the PDMS template is a critical parameter for AI-assisted dry transfer methods, enabling wrinkle-free and crack-free transfer of 2D materials. To quantitatively evaluate this property, atomic force microscopy (AFM) was employed to characterize the template surface morphology with nanometric resolution. A high-precision TESPA-V2 probe was used, and the standard tapping mode of the Bruker Icon system was applied to measure the local surface height variations accurately. As shown in \hyperref[Fig:3target]{Fig.\@ 3(a)}, a representative PDMS template cured at 120 \textcelsius{} exhibits a smooth, dome-shaped profile with a raised center and gently sloping edges. A $1~\mu\mathrm{m}^2$ square region on the top of the dome was scanned to obtain the height distribution map [\hyperref[Fig:3target]{Fig.\@ 3(b)}], which reveals an exceptionally uniform topography. The height variation along a representative central trace (white dashed line in \hyperref[Fig:3target]{Fig.\@ 3(b)}) was extracted and presented in \hyperref[Fig:3target]{Fig.\@ 3(c)}, revealing a peak-to-valley difference of merely 1 nm. The reproducibility of the batch fabrication method was evaluated by statistically analyzing the the root-mean-square (RMS) roughness ($R_q$) values of twenty PDMS templates fabricated at each curing temperature, with curing temperatures systematically varied from 120 \textcelsius{} to 160 \textcelsius{} in 10 \textcelsius{} increments. This gradient-curing approach enables a comprehensive assessment of the surface uniformity and smoothness of the batch-fabricated PDMS templates. The inset histograms in the upper panel of \hyperref[Fig:3target]{Fig.\@ 3(d)} show consistently narrow distributions, with all standard deviations below 0.2 nm. These small variations confirm the excellent surface smoothness, uniformity, and high reproducibility of the developed batch fabrication process. These microscopic roughness features likely arise from temperature-dependent variations in the crosslinking kinetics during the PDMS curing stage. As plotted in \hyperref[Fig:3target]{Fig.\@ 3(d)}, increasing the curing temperature leads to a monotonic rise in RMS roughness. This trend originates from the accelerated crosslinking reactions at elevated temperatures, which promote rapid solidification and limit polymer-chain relaxation, thereby introducing minor nanoscale irregularities. \hyperref[Fig:3target]{Fig.\ 3(e)} compares the surface roughness achieved by \linebreak this batch fabrication method with three literature-reported template-fabrication techniques that employed, photopolymer,\cite{Ref10,Ref11,Ref12,Ref13} resin,\cite{Ref1,Ref2,Ref3,Ref4,Ref5} and metal \cite{Ref6,Ref7,Ref8,Ref9} for template casting. Our method yields markedly lower roughness, with a representative $R_q$ value of approximately 0.3 nm, highlighting its potential for achieving exceptionally smooth surfaces. Consequently, the consistent and reproducible low roughness obtained through batch production ensures the standardized fabrication of transfer stamps, which is crucial for future AI-assisted robotic transfer of 2D materials.

In addition to surface roughness, the geometric uniformity of the PDMS templates represents another critical metric for evaluating the reliability of the batch fabrication process. As illustrated in the inset of \hyperref[fig:4target]{Fig.\@~4(a)}, a representative PDMS template observed under bright-field microscopy (Nikon LV-ND-100) exhibits a well-defined semi-ellipsoidal profile. The corresponding geometric parameters, denoted as $r_a$ (major semi-axis), $r_b$ (minor semi-axis), and $R$ (apex curvature radius), are labeled in the schematic. These parameters, along with the template mass, were statistically analyzed for samples cured at various temperatures. The template masses were determined using a high-precision analytical balance with a 0.1 mg resolution. \hyperref[fig:4target]{Fig.\@ 4(a)} reveals that the apex curvature radius ($R$) exhibits a strong dependence on curing temperature, decreasing significantly from approximately 9 mm to 3 mm as the curing temperature increases from 120 \textcelsius{} to 160 \textcelsius{}. This pronounced trend arises from the accelerated curing and suppressed lateral flow of PDMS at elevated temperatures.\cite{cure1,cure2,cure3} The corresponding histograms fitted with normal distributions (black dashed lines) show a clear narrowing of the distribution width with increasing temperature, from a broad distribution at 120 \textcelsius{}, indicative of higher variability, to a narrow one at higher temperatures, confirming the improved geometric uniformity and reproducibility of the templates. \hyperref[fig:4target]{Fig.\@ 4(b)} compares the evolution of $R$, $r_a$, and $r_b$ across the investigated temperature range. A cooperative relationship between these parameters is observed. With increasing curing temperature, $r_a$ gradually decreases while $r_b$ increases. This coordinated dimensional change gives rise to a sharper semi-ellipsoidal geometry with a more pronounced apex curvature, directly accounting for the observed reduction in $R$. This morphological evolution is consistent with the hypothesis that higher curing temperatures modify PDMS flow dynamics by accelerating solidification, thereby suppressing both lateral spreading (decreasing $r_a$) and gravitational relaxation (increasing $r_b$). The mass of each PDMS template was also measured and statistically analyzed as a function of curing temperature [\hyperref[fig:4target]{Fig.\@ 4(c)}]. The results show a consistent average mass of approximately 14 mg across all curing conditions, with negligible variation as reflected by the narrow error bars. This high reproducibility in mass demonstrates two key aspects: (1) the excellent homogeneity of the PDMS-curing agent mixture achieved by the preparation protocol, and (2) the stable progression of the cross-linking reaction during curing. Overall, the systematic relationship established between the curing temperature and the resulting geometric parameters of the PDMS templates provides a robust foundation for future robotic and AI-assisted transfer of 2D materials. By precisely tuning the curing temperature, one can predefine the curvature radius of the PDMS templates, enabling automated selection of suitable geometries for specific transfer scenarios.

To elucidate the thermomechanical performance of the PDMS templates, a standard 2D material transfer process was implemented, as illustrated in \hyperref[fig:5target]{Fig.\@ 5(a)}. The transfer stamp, consisting of a glass slide, a PDMS template, and a PC film, is mounted on a precision motorized stage with three-axis control. The pickup process involves three steps: motorized descent for initial contact, thermally induced expansion for conformal contact, and cooling-driven pickup. As shown in \hyperref[fig:5target]{Fig.\@ 5(b)}, the descent produces Newton’s rings and then a circular contact zone that expands until reaching the edge of the target 2D material ($\sim 200~\mu\mathrm{m}$ from the center). Subsequent substrate heating promotes expansion of the contact front beyond the material boundary, while controlled cooling triggers contraction-induced pickup. This temperature-mediated mechanism ensures gentle detachment without mechanical fracture, underscoring the importance of characterizing PDMS thermal expansion. The substrate temperature was precisely controlled and monitored by a Pt100-equipped temperature controller, cycling from 40 \textcelsius{} to 150 \textcelsius{} (15 min hold) and back to 40 \textcelsius{} at ± 0.05 \textcelsius/s [\hyperref[fig:5target]{Fig.\@ 5(c)}]. The measured temperature closely followed the programmed profile, verifying thermal stability. Heating induced contact-area expansion, whereas cooling caused contraction for material pickup. The evolution of the contact radius was extracted using an image-processing algorithm and plotted against the substrate temperature ($T_s$) in \hyperref[fig:5target]{Fig.\@ 5(d)}. Five replicate templates were analyzed for each curing temperature, and logistic fits were applied to describe the polymer expansion.\cite{RN126,RN130,RN131,RN132} As shown in \hyperref[fig:5target]{Fig.\@ 5(e)}, PDMS thermal expansion depends strongly on curing temperature. Higher curing temperatures yield slower expansion, with total displacements decreasing from approximately 800 $\mu$m (120 \textcelsius{}) to 600 $\mu$m (160 \textcelsius{}). This effect arises from two curvature-related factors. First, templates with smaller apex radii (formed at higher temperatures) require greater compressive stress for conformal contact, limiting subsequent expansion. Second, the increased curvature results in a larger contact angle with the substrate, geometrically reducing the effective expansion area. Together, these effects suppress expansion at higher curing temperatures. This systematic thermomechanical characterization yields a set of reproducible logistic expansion curves for each curing temperature, forming a quantitative database for predictive control. Such data are essential for optimizing future AI-assisted robotic transfer processes, where template geometry and thermal response can be precisely tuned to different 2D material systems.

In conclusion, we have developed a batch fabrication strategy for PDMS templates with highly uniform geometry and well-defined thermomechanical responses, tailored for robotic transfer of 2D materials. Within the explored curing-temperature range (120 - 160 \textcelsius{}), lower curing temperatures produce smoother template surfaces with RMS roughness down to 0.3 nm, benefiting from slower cross-linking and enhanced surface relaxation. In contrast, higher curing temperatures yield templates with smaller apex curvature radii and slower, more stable thermal-expansion dynamics. The latter leads to a gradual evolution of Newton’s rings during heating, enabling accurate prediction and control of the radial displacement of the PC and substrate contact boundary, an essential feature for AI-assisted transfer precision. By quantitatively correlating curing parameters with surface morphology, geometric dimensions, and thermal expansion characteristics, this study establishes a reproducible parametric database for PDMS template engineering. Such a standardized and data-driven foundation is expected to play a pivotal role in the development of intelligent, feedback-controlled 2D material transfer systems.


\section*{Author Declarations}

\noindent\textbf{ Conflict of Interest}

The authors have no conflicts to disclose.

\begin{acknowledgments}
This work is supported by the National Natural Science Foundation of China (NSFC) (Grant Nos. 62204145, 12204287, 12374185, 62274180 and U21A6004). B.D. acknowledges the support of the Fund for Shanxi “1331 Project” Key Subjects Construction. B.D acknowledge supports from the Quantum Science and Technology-National Science and Technology Major Project (Grant No. 2021ZD0302003).
\end{acknowledgments}
\section*{Data Availability Statement}

The data that support the findings of this study are available from
the corresponding author upon reasonable request.

\section*{REFERENCES}
\bibliography{aipsamp}

\end{document}